\documentclass[%
 reprint,
preprintnumbers,
nofootinbib,
nobibnotes,
 amsmath,amssymb,
 floatfix
 prd
]{revtex4-1}
\usepackage{setspace}
\usepackage{times}
\usepackage{CJK}
\usepackage{lipsum}
\usepackage{float}
\usepackage{graphicx}% Include figure files
\usepackage{dcolumn}% Align table columns on decimal point
\usepackage{bm}% bold math
\usepackage[colorlinks=true,pdfstartview=FitV,breaklinks=true]{hyperref}
\usepackage[dvipsnames,table]{xcolor}
\hypersetup{urlcolor=Black, citecolor=Black, linkcolor=Black}
\begin{document}

\preprint{CPPC-2020-03}

\title{Topologically induced black hole charge and its astrophysical manifestations}

\author{Yunho Kim}
 \email{yunho.kim@sydney.edu.au}
\author{Archil Kobakhidze}
 \email{archil.kobakhidze@sydney.edu.au}
\affiliation{Sydney Consortium for Particle Physics and Cosmology, \\
 School of Physics, The University of Sydney, NSW 2006, Australia 
}

\begin{abstract}
Quantum corrected effective action for gravity contains massive spin-2 ghost degrees of freedom and admits a topological term which couples longitudinal vector degrees of freedom of the massive spin-2 to Maxwell's electromagnetism. We argue that in the presence of otherwise unobservable gravimagnetic poles this topological term induces an electric charge on a black hole which can be probed through the associated electric and magnetic fields. In particular, we discuss the electromagnetic follow up from the LIGO-sensitive charged black hole binary coalescence due to the synchrotron radiation from the surrounding plasma and the shadow of super-massive charged black holes.
\end{abstract}

\maketitle
%\tableofcontents

\section{Introduction}
The quantum nature of matter inevitably implies modification of the Einstein theory of General Relativity (GR). The radiative effects of virtual particles lead to the higher powers of the Riemann and Ricci curvature terms to be induced in addition to the Einstein-Hilbert term in the quantum corrected effective action for gravity \cite{Utiyama:1962sn}. These corrections embody new degrees of freedom describing massive scalar and massive ghost spin-2 particles in the effective low-energy description. Within string theory, such modification of GR is expected to be significant only at minuscule Planckian scales \cite{Alvarez-Gaume:2015rwa} and hence is unlikely to be observable by any classical measurement in the foreseeable future. Within the semiclassical gravity approach, the effects are not very prominent either and the constraints from the available experimental data are rather mild \cite{Kim:2019sqk}.

In Ref. \cite{Dvali:2006az}, it has been argued that the massive spin-2 field may modify the GR black hole solution by inducing a quantum hair \cite{Coleman:1991ku}. This type of hair is not detectable by classical measurements based on local interactions, but rather through a quantum interference effect. Subsequently, one of us has shown that classically unobservable gravimagnetic monopole hair of the massive spin-2 field could couple to the electromagnetic field via a topological term. Such interactions then induces an electric charge onto a black hole which can be detected by measuring the associated long range electric field. This effect is reminiscent of the Witten effect \cite{Witten:1979ey}, a magnetic monopole acquiring an electric charge in the presence of a topological electromagnetic theta-term. The crucial difference though is that the monopole gravimagnetic field is unobservable, unlike the field produced by the magnetic monopole. In this work we would like to further explore this topological mechanism for black hole charging and investigate potential astrophysical manifestations of this phenomenon.

According to the black hole no-hair conjecture \cite{Ruffini:1971bza}, a generic black hole carries only mass, angular momentum and electric charge. For astrophysical black holes, however, the electric charge is believed to be negligibly small, since no efficient mechanism for charging is known. Indeed, if the black hole acquires its charge by the accretion of charged matter, the accretion stops whenever the attractive Newtonian force is counterbalanced by the repulsive Coulomb force. This results in the limit on the accreted charge $Q$:
\begin{equation} \label{charge1}
    \mathfrak{q} \equiv\frac{|Q|}{\sqrt{G}M}\leq \frac{m}{|q|M_P}\sim 2.6\cdot 10^{-19}\left(\frac{m}{1~{\rm GeV}}\right)\left(\frac{e}{|q|}\right)~,
\end{equation}
where $M$ is the black hole mass, $m$ and $q$ are the mass and the charge of the accreted matter particles, $e=\sqrt{4\pi\alpha}\approx 0.303$ and $M_P\equiv 1/\sqrt{G}\approx 1.2\cdot 10^{19}$ GeV is the Planck mass. In the above numerical estimation we have assumed the accreted particles to be protons, $m=m_p\approx 1$ GeV, $q=e$. Alternatively, if a charged black hole were formed by the collapse of matter with a sizeable charge, it undergoes a rapid discharge due to the conductive ambient plasma. Conservatively, the charge accretion rate is given by the Eddington formulae, $\frac{d|Q|}{dt}=\frac{32\pi}{3}\epsilon\frac{GMm_p^3}{\alpha^2}{\frac{|q|}{m}}$ [for the Eddington accretion the radiativity is $\epsilon =1/10$]. Thus, the discharge time,
\begin{equation} \label{discharge}
    \tau \simeq 4.3\cdot 10^{-10}\left(\frac{m}{m_p}\right)\left(\frac{e}{|q|}\right)\left(\frac{\mathfrak{q}}{1}\right)~s~,
\end{equation}
is extremely fast even for the extremal black holes ($\mathfrak{q}=1$) when the elementary charge carrier particles are assumed to be protons ($m=m_p, q=e$). This constraint can be somewhat relaxed if plasma around the black hole is populated by a stable massive particles carrying very small electric charge, see Ref. \cite{Cardoso:2016olt}.

Finally, the equilibrium neutral plasma conspiring of protons and electrons around the black hole gets polarized due to the proton-electron mass difference \cite{Eddington, Harrison}. The accumulated (negative) charge on the black hole is again very small:
\begin{equation} \label{charge2}
    \mathfrak{q}=\frac{m_p-m_e}{2e M_P}\simeq 1.4\cdot 10^{-19}~.
\end{equation}

The above considerations imply that the electric charge of astrophysical black holes are constrained to be tiny, $\mathfrak{q}\lesssim 10^{-19}$. Such a small charge cannot manifest in any significant way in astrophysical observations and hence astrophysical black holes are commonly regarded as being electrically neutral.

None of the above-discussed constraints, however, are applicable to the topologically charged black holes proposed in this paper, since the topological interactions are unaffected by the local dynamics. Therefore, \emph{a priori} large topologically induced electric charge ($\mathfrak{q}\sim \mathcal{O}(1)$) could actually be carried by black holes. This phenomenon then may manifest in astrophysical observations as will be discussed below.

The rest of the paper is organised as follows. In the next section we describe the topological charging of black holes within the so-called black hole membrane paradigm \cite{Thorne:1986iy, Parikh:1997ma}. We explicitly solve for the black hole spacetime by carefully choosing the membrane boundary term that accounts for the hidden gravimagnetic charge. We confirm that the solution formally is the Reissner–Nordström spacetime, but with non-zero electric charge homogeneously distributed near the black hole horizon. This charge is entirely determined by the coupling of the topological term and the otherwise unobservable gravimagnetic charge. In sec. \ref{astro} we discuss two potential astrophysical manifestations of the induced electric charge. The first is the unconventional dynamics of charged black hole binaries and the second is the modified black hole shadow caused by the electric charge. The final sec. \ref{concl} is reserved for conclusions.

\section{Induced electric charge via a topological term} \label{bh_charge}
\subsection{Quadratic gravity, massive spin-2 and topological interactions with electromagnetism}
Let us start by recalling the basics of quadratic gravity. The most general diffeomorphism-invariant action of gravity with up to the quadratic in curvature terms has the form \footnote{A possible term proportional to the square of the Riemann tensor, $R_{\mu\nu\rho\sigma}R^{\mu\nu\rho\sigma}$, can be eliminated using the Gauss-Bonnet identity.}:
\begin{equation} \label{oriact1}
    S = \int d^4x \sqrt{-g} \left[\frac{R}{16\pi G} + \beta R^2 + \gamma R^{\mu \nu} R_{\mu \nu} \right],
\end{equation}
where $\beta$ and $\gamma$ are the dimensionless parameters. The quadratic in curvature terms in (\ref{oriact1}) embody extra propagating degrees of freedom beyond the massless graviton of GR attributed to the massive scalar and the massive spin-2 fields. In this paper we are not interested in the additional scalar degree of freedom and to remove it we impose the condition: $4\beta=-\gamma$. To make the massive spin-2 degrees of freedom manifest we introduce an auxiliary tensor field $\pi_{\mu\nu}$ and rewrite the action (\ref{oriact1}) in an equivalent form:   
\begin{equation} \label{oriact2}
    S = \int d^4x \sqrt{-g} \left[\frac{R}{16\pi G} + \pi^{\mu \nu} \Pi_{\mu \nu} -\frac{1}{4\beta} \pi^{\mu \nu} \pi_{\mu \nu} \right],
\end{equation}
where $\Pi_{\mu \nu}=R_{\mu \nu}-\frac{1}{4}g_{\mu \nu}R$. The Euler-Lagrange equation for $\pi_{\mu\nu}$ obtained from (\ref{oriact2}) implies $\pi_{\mu \nu}=2\beta\Pi_{\mu \nu}$. This substituted back into (\ref{oriact2}) gives the original action (\ref{oriact1}) with $\gamma = -4\beta$. This alternative form of the action for quadratic gravity makes it immediately clear that the vacuum black hole solutions ($R_{\mu\nu}=R=0$) of GR, such as the Schwarzschild solution, are also automatically solutions of quadratic gravity with $\pi_{\mu\nu}=0$. 

We would like to extend the above action (\ref{oriact2}) by adding a topological term which couples quadratic gravity to Maxwell's electromagnetism. To this end let us decompose the traceless tensor field $\pi_{\mu\nu}$ into a divergenceless and traceless tensor $\tilde \pi_{\mu\nu}$ ($\partial^{\mu}\tilde \pi_{\mu\nu}=0$) field and a 4-vector $B_{\mu}$  field:
\begin{equation} \label{pi1}
    \pi_{\mu \nu} = \tilde{\pi}_{\mu \nu} + \partial_{\mu} B_{\nu} + \partial_{\nu} B_{\mu}.
\end{equation}
The topological term we are after then has the form:
\begin{equation} \label{top}
    S_{\mathrm{top}} = \int d^4x ~\alpha\epsilon^{\mu\nu\rho\sigma}F_{\mu\nu}B_{\rho\sigma}~,
\end{equation}
where $F_{\mu\nu}=\partial_{\mu}A_{\nu}-\partial_{\nu}A_{\mu}$ is the electromagnetic field strength, $B_{\mu\nu}=\partial_{\mu}B_{\nu}-\partial_{\nu}B_{\mu}$ and $\alpha$ is a coupling constant. Note that the topological term (\ref{top}) does not contain the metric tensor and does not influence the dynamics of local dynamical degrees of freedom. However, it can modify global solutions of quadratic gravity, the Schwarzschild solution in particular, as is shown below.

\subsection{Membrane paradigm and topologically induced electric charge}
The existence of an  event horizon is a defining property of black holes. Mathematically it is defined as a (2+1)-dimensional null boundary hypersurface inside which light cones tip over and time coordinate becomes spatial. However, physically it is undetectable by means of any local measurement and the description of physical phenomena often have trouble accommodating the event horizon. Therefore, for astrophysical applications the membrane paradigm has been developed \cite{Thorne:1986iy}, where instead of the event horizon a (2+1)-dimensional time-like hypersurface (stretched horizon) located just outside the event horizon is considered. The stretched horizon has a non-singular induced metric and thus provides a more tractable boundary for external fields. Furthermore, the equations governing the stretched horizon represent a very good approximation to those for the true horizon \cite{Thorne:1986iy}. We adopt here this treatment of black holes and follow the formalism of Ref. \cite{Parikh:1997ma}.

The formalism refers to the physics relevant for an observer outside the black hole horizon. The corresponding dynamics is thus determined by varying the part of the total action restricted to the spacetime outside the black hole, $S_{\mathrm{out}}$. This external action, however, is not stationary by itself unless boundary conditions are fixed at the stretched horizon. Hence, to obtain the correct equations of motion $S_{\mathrm{out}}$ must be augmented by boundary terms defined at the stretched horizon (in addition to the familiar Gibbons-Hawking term at infinity, which is not explicitly shown here):
\begin{equation} \label{membrane1}
    S=S_{\mathrm{out}}+S_{\mathrm{boundary}}~. 
\end{equation}
For the problem at hand we have:
\begin{eqnarray} \label{out}
    S_{\mathrm{out}}=\int d^4x \sqrt{-g} \left[\frac{R}{16\pi G} -\frac{1}{4}F_{\mu\nu}F^{\mu\nu}+\right. \nonumber \\
    \left. \frac{\alpha}{\sqrt{-g}} \epsilon^{\mu\nu\rho\sigma}F_{\mu\nu}B_{\rho\sigma}\right]~,
\end{eqnarray}
\begin{equation} \label{bound}
    S_{\mathrm{boundary}}=\int d^4x \sqrt{-g}\delta(r-r_{h})J^{\mu}A_{\mu}~,
\end{equation}
where $r=r_h$ is the position of the stretched horizon and the boundary current reads as
\begin{equation} \label{current1}
    J^{\mu}=F^{\mu\nu}n_{\nu}-4\alpha \epsilon^{\mu\nu\rho\sigma}B_{\nu\rho}n_{\sigma}
\end{equation}
[$n_{\mu}$ is an unit vector normal to the stretched horizon]. In the above equations (\ref{out}) and (\ref{bound}) we have assumed $\pi_{\mu\nu}=0$ in accord with the vacuum black hole solution in quadratic gravity. Triviality of spin-2 tensor field $\pi_{\mu\nu}$ does not exclude, however, the non-trivial gravimagnetic solution:
\begin{equation} \label{current2}
    B_{\theta \phi}=-\frac{q_{gm}}{2}\sin\theta~,
\end{equation}
where $q_{gm}$ is the charge of the gravimagnetic pole. This locally unobservable gravimagnetic pole induces electric charge on the \emph{ab initio} neutral black hole via the `secrete' topological interaction (\ref{top}). Indeed, this can verified explicitly by adopting the spherically symmetric ansatz for the metric,
\begin{equation} \label{symmetric}
    ds^2 = A(t,r)dt^2 - B(t,r)dr^2 - r^2 d\theta^2 - r^2 \sin{\theta}d\phi^2
\end{equation}
and solving the equations of motion obtained from the field variations of the action (\ref{membrane1}). The relevant equations followed from the variation with respect to the electromagnetic 4-vector potential $A_{\mu}$ are:
\begin{align}
    & \partial_{0} F^{10} = 0, \label{ele1} \\
    & \partial_{1} F^{01} + F^{01} \partial_{1} \ln{\left(\sqrt{AB} r^2 \right)} = \frac{Q \delta(r-r_h)}{4 \pi r^2}~, \label{ele2}
\end{align}
where $Q\equiv q_m\alpha$. The first Eq. (\ref{ele1}) implies that the radial component of the electric field is static, while Eq. (\ref{ele2}) gives the solution:
\begin{equation}
    F^{01} = \frac{C}{\sqrt{AB} r^2} + \frac{Q}{4\pi r^2}\left(\frac{\left. AB\right |_{r=r_h}}{AB}\right)^{1/2}~,
\end{equation}
where $C$ is the integration constant, which we set $C=0$, since the black hole without the topological term is assumed to be electrically neutral. The electromagnetic scalar potential then reads:
\begin{equation} \label{sol1}
    A_{0} = \frac{Q}{4\pi r}\left(\frac{\left. AB\right |_{r=r_h}}{AB}\right)^{1/2}~,
\end{equation}
and the vector potential $\vec A=0$.

The explicit form of the metric functions $A$ and $B$ is obtained from the Einstein equations:
\begin{align}
    R_{\mu \nu}-\frac{1}{2}g_{\mu \nu}R = \kappa \left[T^E_{\mu\nu} + T^B_{\mu\nu} \right]~, \label{eom1}
\end{align}
where $T^E_{\mu\nu}=\frac{1}{4}g_{\mu\nu}F^{\alpha \beta}F_{\alpha \beta}-F^{\mu \alpha} F_{\nu \alpha}$ is the standard electromagnetic energy-momentum tensor and $T^B_{\mu\nu}= \left(J_{\mu}A_{\nu}+J_{\nu}A_{\mu}-g_{\mu\nu}J_{\alpha}A^{\alpha} \right) \delta(r-r_h)$ is the boundary energy-momentum tensor. Namely, we find: 
\begin{equation} \label{sol2}
    A =1/B= 1 + \frac{D}{r} + \frac{G Q^2}{r^2}~,
\end{equation}
where $B=1/A$. By fixing the integration constant $D=-2GM$ through the black hole mass $M$, we obtain the Reissner–Nordström metric for the black hole exterior. The key difference with the electro-vacuum Reissner–Nordström solution is that the topologically induced black hole charge $Q= q_{gm}\alpha$ is distributed on the stretched horizon, just outside the black hole. As for the Reissner–Nordström black hole, we require the induce charge
\begin{equation} \label{topcharge}
    \mathfrak{q}\equiv \frac{Q}{\sqrt{G}M}\leq 1~,
\end{equation}
where the inequality is saturated by the extremal black hole solution. The topologically induced charge (\ref{topcharge}) is entirely defined by the charge of a gravimagnetic pole $q_{gm}$ and the topological coupling $\alpha$, and can be much larger than charge accumulated due to the local interactions, see Eqs. (\ref{charge1}) and (\ref{charge2}). Furthermore, as this charge is sourced by a classically unobservable \cite{Dvali:2006az} gravimagnetic pole via topological interactions (\ref{top}), no local dynamical process can affect it. Therefore, we would like to conclude that stable black holes with a large electric charge (\ref{topcharge}) can theoretically exist. Astrophysical black holes with a significant electric charge may have interesting observational manifestations which we discuss in the next section.

\section{Astrophysical manifestations of the induced electric charge} \label{astro}
\subsection{Electromagnetic radiation from the charged black hole binary coalescence}
Up until recent times, studies of charged astrophysical black holes attracted a limited number of attention because of the common belief that black holes in astrophysical setting are essentially neutral. After the discovery of gravitational waves from the binary black hole coalescence, there was an increased interest in charged black holes, largely motivated by the initial reports \cite{Connaughton:2016umz} on the apparent electromagnetic counterpart of the mergers of two black holes. In particular, it has been pointed out in \cite{Zhang:2016rli} that the magnetic dipole radiation could account for the apparent short gamma-ray bursts (GRBs) accompanying the merger of two black holes if at least one of the black holes carries the electric charge $\mathfrak{q}\sim 10^{-5}-10^{-4}$, while smaller charges $\mathfrak{q}\sim 10^{-9}-10^{-8}$ can explain the origin of the mysterious fast radio bursts (FRBs) \cite{Lorimer:2007qn}. In \cite{Liu:2016olx} the authors suggested that rotating black holes with a large enough charge can be responsible for FRBs. Subsequently, it has been noted in \cite{Deng:2018wmy} that for charged black hole binaries, dominant radiation may come in the form of electric dipole radiation. Other recent related studies include Refs. \cite{Liebling:2016orx, Fraschetti:2016bpm, Levin:2018mzg, Zhang:2019dpy, Cardoso:2016olt, Wang:2020fra}. These studies show that a sizeable black hole charge, exceeding the standard limits (\ref{charge1}, \ref{charge2}) by several orders of magnitude, is required to be responsible for the observable electromagnetic follow up. Such large black hole charges can be thought as being induced through our topological mechanism.

Some of the previous studies, however, were focused on charged black hole coalescence in empty space, while the astrophysical black holes are actually immersed in the interstellar plasma. In fact, the interstellar plasma frequency $\omega_{\rm plasma}\approx \sqrt{e^2 n_e/m_e}\simeq 18~\mathrm{kHz}\left(n_e/0.1\mathrm{cm}^{-3}\right)^{1/2}$ is an order of magnitude larger than the anticipated frequency of the electromagnetic radiation from the coalescence of charged black holes with masses in the LIGO range $\sim 10-30~\mathrm{M}_{\odot}$. The interstellar plasma, therefore, is not transparent for such low frequency radiation. Furthermore, the Debye length of the interstellar plasma, $\lambda_D=\sqrt{T_e/e^2 n_e}\sim \mathcal{O}(10\textrm{m})$, is much smaller than the typical distance between binary black holes ($\gtrsim 100\textrm{km}$). This implies that the electrostatic Coulomb potential between charged black holes is Debye-screened by the surrounding plasma and thus does not contribute significantly to the dynamics of binaries and their electric dipole radiation must be suppressed.

Here we make estimations about the synchrotron electromagnetic radiation of plasma electrons moving in the magnetic field of coalescing charged black holes. To this end, let us consider two black holes with masses $m_{1,2}$ and topologically induced electric charges $q_{1,2}$ which form a binary system. Since these black holes are charged and orbit each other they carry magnetic moments,
\begin{equation} \label{magnetic1}
    \mathfrak{m}_{1,2}=\frac{q_{1,2}}{2m^2_{1,2}}\mu^2\sqrt{G Ma}=\frac{\mathfrak{q}_{1,2}}{2m_{1,2}}\mu^2\sqrt{G^2 Ma}~,
\end{equation}
which are directed perpendicular to the coalescence plane. In Eq. (\ref{magnetic1}) $M=m_1+m_2$ and $\mu=m_1m_2/M$ is the total and the reduced masses of the binary system respectively, and $a$ is the distance between black holes. Note that we have ignored here the electrostatic Coulomb interactions between charged black holes, since as has been pointed out earlier these interactions are Debye-suppressed in the surrounding plasma. Therefore, the orbital period $T=2\pi a^{3/2}(GM)^{-1/2}$ (in the Newtonian circular orbit approximation) and the orbital decay is determined by the attractive gravitational force. Since the photon is effectively massive in plasma, the generated magnetic field is confined in the narrow region around the orbit. We can then estimate the magnitude of the magnetic field of the binary as:
\begin{equation} \label{magnetic2}
    B=\frac{|\mathfrak{m}_1+\mathfrak{m}_2|}{4\pi a^3}=
    \frac{G\mu^2M^{1/2}}{8\pi a^{5/2}}
    \left( \frac{\mathfrak{q}_1}{m_1}+\frac{\mathfrak{q}_2}{m_2}\right)
\end{equation} 
This magnetic field can indeed be very strong. E.g., for $m_1=m_2=30 M_{\odot}$, $\mathfrak{q}_1=\mathfrak{q}_2=1$ the magnitude of the magnetic field at $a\approx GM$ is $B\approx 1.4\cdot 10^{12}~\textrm{T}$ which is of some three orders of magnitude stronger than the quantum critical magnetic field $B_c=m_e^2/e\approx 4.41\cdot 10^9~\textrm{T}$ and comparable with the typical magnetic field of magnetars. Note that for binaries of oppositely charged black holes with $\mathfrak{q}_1m_2=-\mathfrak{q}_2m_1$ the magnetic field (\ref{magnetic2}) gets cancelled out.

The charged particles (electrons) of the interstellar plasma would move on helical trajectories with relativistic velocities around the magnetic field and emit synchrotron electromagnetic radiation. Most of the synchrotron radiation is emitted at the peak critical frequency (at the pitch angle $\alpha=\pi/2$) which is equal to\footnote{In this estimation we ignore time variation of the induced magnetic field, since the rotational frequency of plasma electrons are much higher than the frequency of the binary coalescence.}:
\begin{align}
    \omega_{c} & \simeq \frac{3}{2}\gamma^2_{\rm max}m_e\frac{B}{B_c} \nonumber \\
    & \approx 2.6\cdot 10^{13}\left(\frac{\gamma_{\rm max}}{10^7}\right)^2\left(\frac{B}{10^{12}\textrm{T}}\right)~\textrm{YHz}~.
\end{align}
The charged black hole binary luminosity (averaged over the pitch angle) is:  
\begin{eqnarray} \label{lumi}
    &&L_{\rm CBHB} \simeq \frac{1}{9 \pi}N_e\gamma_{\rm max}^2m_e^2\left(\frac{B}{B_c}\right)^2 \approx \nonumber \\
    && 9.8\cdot 10^{47}\left(\frac{\gamma_{\rm max}}{10^7}\right)^2\left(\frac{B}{10^{12}\textrm{T}}\right)^2\left(\frac{N_e}{8.5\cdot 10^{15}}\right)~\textrm{erg/s},
\end{eqnarray}
where we used a somewhat naive estimation of the number of electrons in magnetosphere: $N_e\sim n_e a^2\lambda_D\approx 8.5\cdot 10^{15}$. Hence, maximally charged black hole binary in the LIGO mass range can be a very bright source of ultra-high energy gamma rays.

Nevertheless, detection of synchrotron radiation from the black hole binaries as the follow up to the associated gravitational waves is quite a challenging task. The current sensitivity of detectors seems to impose the following limit on the energy of electromagnetic follow up: $E\lesssim 10^{49}~\textrm{erg}$ \cite{Perna:2019pzr}. Our estimated luminosity (\ref{lumi}) gives a much smaller total radiated energy. So no limits on the charge of black holes can be set at present. In this regard, it would be perhaps interesting to reverse the process and, e.g., search for gravitational follow up of short GRBs.

It is indeed rather intriguing to speculate that charged black hole binaries are sources of mysterious short GRBs and FRBs \cite{Zhang:2016rli}. Our estimation of the luminosity Eq. (\ref{lumi}) falls short of the typical short GRB luminosity $\sim 10^{51}~\textrm{erg/s}$. The synchrotron radiation cannot account for FRBs either. Perhaps other considerations \cite{Zhang:2016rli} and \cite{Liu:2016olx} aided by our mechanism of charge generation are more promising in this regard.

The black hole charge also modifies gravitational waveforms and in principle can be inferred/constrained from the precise measurement of the phase of gravitational waves \cite{Cardoso:2016olt, Wang:2020fra}. However, since the long-range Coulomb interactions of black holes are suppressed in the interstellar medium, we expect that the corresponding modification of waveforms will be very challenging to observe  with current detectors.

\subsection{Shadow of the charged black hole}
Photons passing a black hole nearby the horizon are strongly influenced by the black hole geometry. In particular, the black hole charge may substantially modify the motion of photons at small radii, while the charge effect is negligible at large radial coordinates of photon trajectories.

The critical impact parameter for non-rotating charged black hole with charge $\mathfrak{q}$ can be computed as \cite{Zakharov:2014lqa}:
\begin{equation} \label{impact}
    l_{cr}=\frac{8\mathfrak{q}^4-36\mathfrak{q}^{2}+27+\sqrt{-512(\mathfrak{q}^{2}-9/8)^3}}{2(1-\mathfrak{q}^{2})}~.
\end{equation}
Note, we require that the black hole charge does not exceed its extremal value (\ref{topcharge}), which automatically avoids the issue of unstable photon orbits occurring for $\mathfrak{q}>3/2\sqrt{2}$. The black hole shadow size then reads:
\begin{equation} \label{size}
    D_{shad}=l_{cr}^{1/2}\frac{R_{S}}{d}~,
\end{equation}
where $R_S=2GM_{bh}$ is the Schwarzschild radius of the black hole of mass $M_{bh}$ and $d$ is the distance to the black hole. One can observe that the size of the charged black hole is smaller than the charge of the same mass neutral black hole.

\begin{figure}[t]
\includegraphics[width=0.5\textwidth]{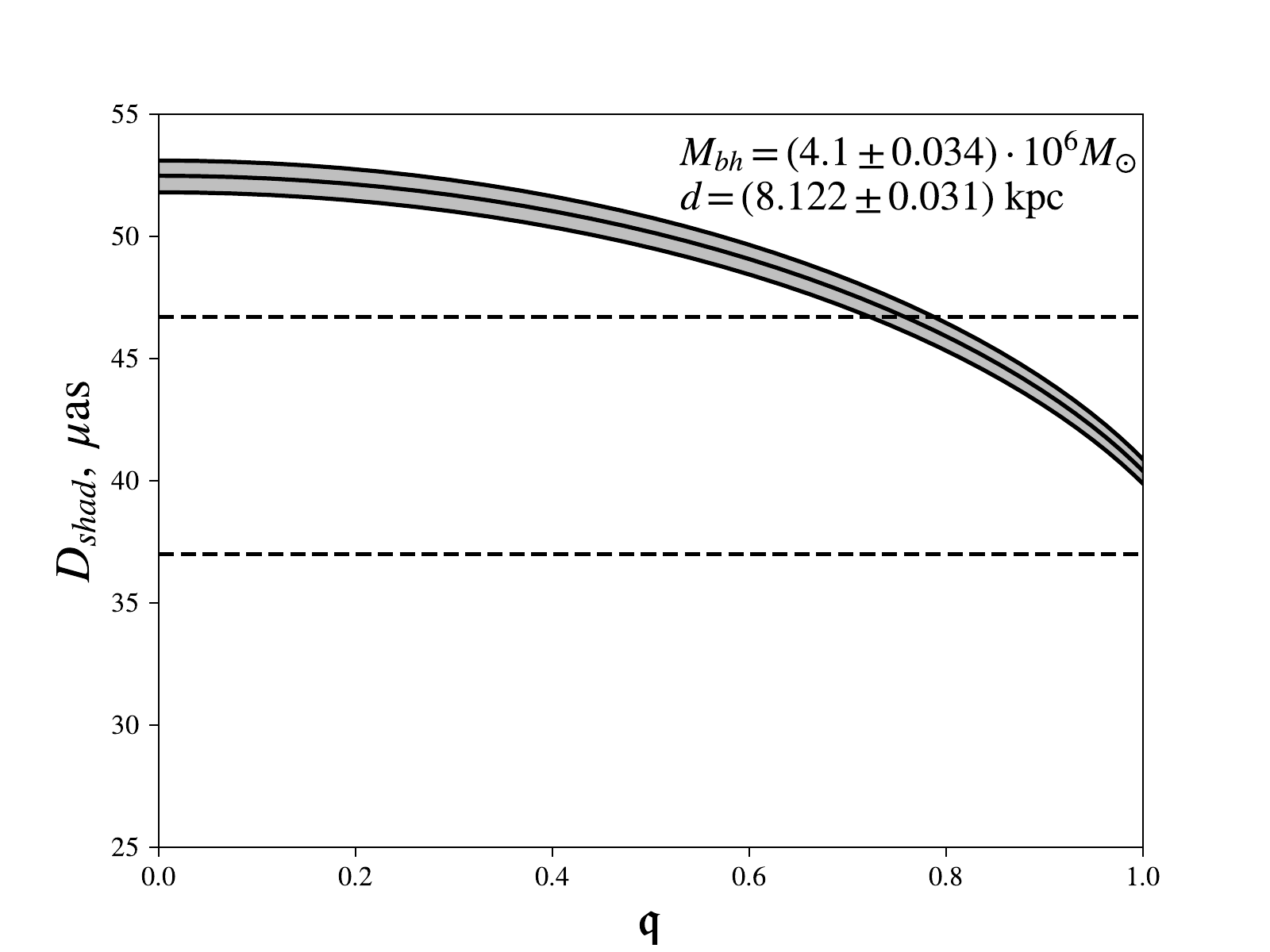} \\
\includegraphics[width=0.5\textwidth]{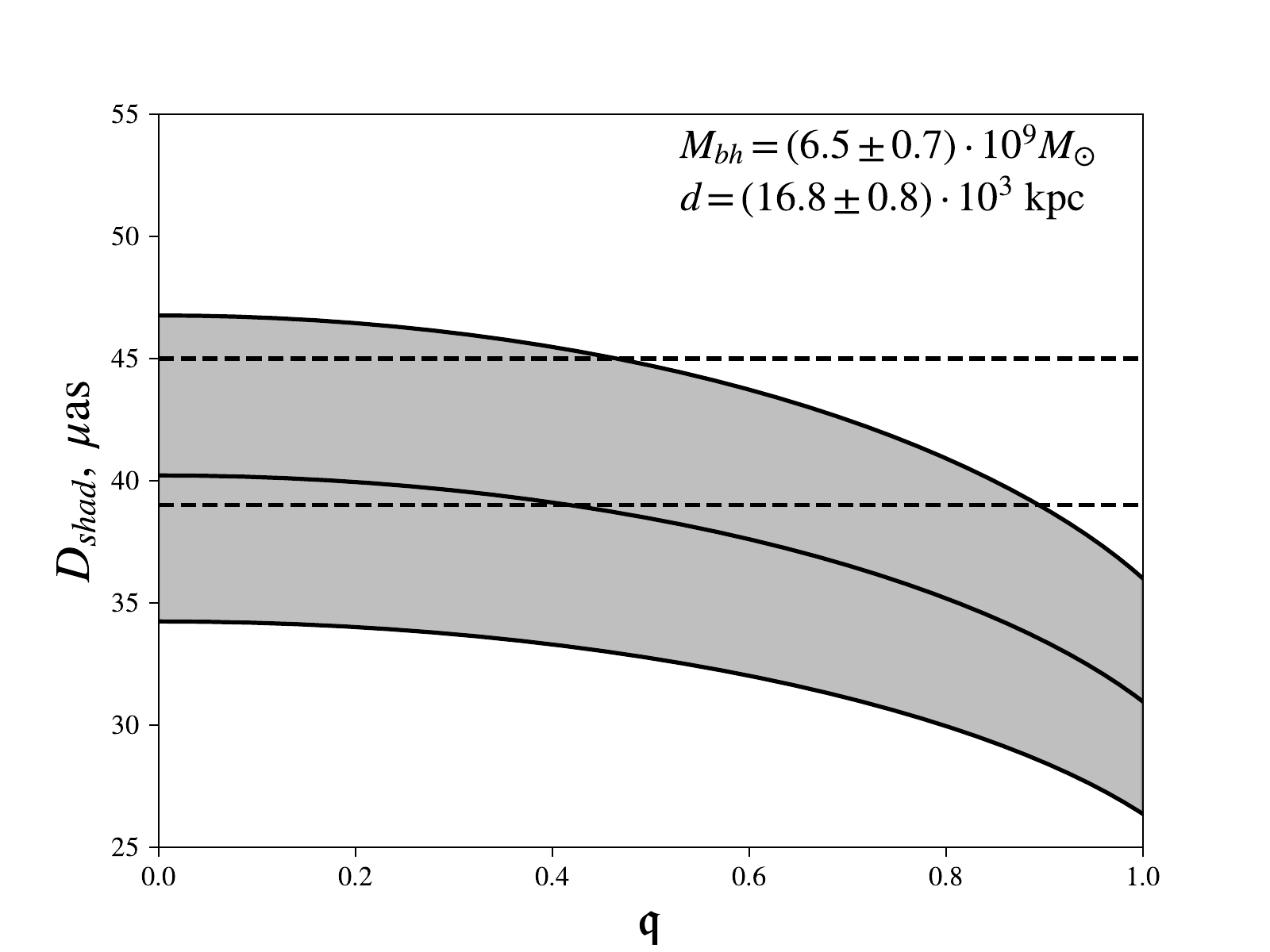}\caption{\label{Figure} The shadow size $D_{shad}$ vs black hole charge $\mathfrak{q}$ (shaded regions) for the Galactic Centre (upper panel) and M87 (lower panel) black holes. The horizontal bands bounded by the dashed lines on each panel represent the measured sizes of the Galactic Centre \cite{Fish} and M87 \cite{Akiyama:2019cqa} black holes.}
\end{figure}

In Fig. \ref{Figure} we have plotted the shadow size as a function of charge for the Galactic Centre black hole (shaded band in the upper panel) and M87 black hole (shaded band in the lower panel). The masses and distance to those black holes are $(M_{bh}=(4.1\pm 0.034)\cdot 10^6 M_{\odot},~d=(8.122\pm0.031)~\textrm{kpc})$ \cite{Abuter:2018drb} and $(M_{bh}=(6.5\pm 0.7)\cdot 10^9 M_{\odot},~d=(16.8\pm0.8)\cdot 10^3 ~\textrm{kpc})$ \cite{Akiyama:2019cqa}, respectively. The most accurate measurement of the size of the Galactic Centre gives $D_{shad}= 41.3^{+5.4}_{-4.3}~\mu\textrm{as}$ (at $3\sigma$ C.L., on day 95) \cite{Fish}. These experimental values are shown as the band between horizontal dashed line in the top panel of Fig.  \ref{Figure}. Intriguingly, the data favours a black hole with the significant near extremal charge, $\mathfrak{q}\gtrsim 0.8$ (see, also \cite{Zakharov:2014lqa}). The expected more accurate measurements by the Event Horizon Telescope (EHT) \cite{Goddi:2017pfy} may confirm (or falsify) this preliminary finding.

Recently, the EHT Collaboration have also measured the size of M87 black hole, $D_{shad}= 42\pm 3~\mu\textrm{as}$ \cite{Akiyama:2019cqa}. The experimentally allowed values are shown as the band bounded by the horizontal dashed lines on the lower panel of Fig. \ref{Figure}. While the date are consistent with the neutral black hole, constraints on the charge are very weak, still allowing large charges   $\mathfrak{q}\lesssim 0.5-0.9$.

%%%%%%%%%%%FIGURE%%%%%%%%%%%%%%%%%%%%%%%%
%%%%%%%%%%%%%%%%%%%%%%%%%%%%%%%%%%%%%%%

\section{Conclusions} \label{concl}
In this paper we have proposed a topological mechanism for charging astrophysical black holes within GR with extra quadratic in curvature terms generically induced through the effects of quantum matter. We explicitly demonstrate that an electric charge is induced on the black hole in the presence of classically unobservable gravimagnetic pole of the effective massive spin-2 field that interacts with electromagnetism via the topological interaction (\ref{top}). Due to the topological origin of the induced charge, it cannot be neutralised by the standard local electromagnetic processes and is only theoretically bounded by its extremal value.

We have then discussed coalescence of charged black hole binaries. For binaries in LIGO-mass range immersed in the interstellar medium, the dipole electromagnetic radiation cannot propagate in plasma and, furthermore, the electromagnetic interactions between two black holes is Debye-screened. The charged binaries are, however, strong sources of very high energy synchrotron radiation, and in fact could power short GRBs. Our estimate indicates, however, that the sensitivity of current detectors is not enough to identify the electromagnetic radiation from LIGO-mass charged black hole binaries as the follow up of the associated gravitational waves. With the increased accuracy of future multi-messenger observations, one should be able to establish whether the binaries carry the net electric charge or not.

We have also considered the effect of the electric charge on shadows of supermassive black holes. Intriguingly, current data favour very large $\mathfrak{q}\gtrsim 0.8$ charge for the Galactic Centre black hole. The measured shadow size for the M87 black hole is consistent with its electric neutrality, although the current constraint on the charge is rather weak $\mathfrak{q}\lesssim 0.5-0.9$.

In summary, our topological mechanism for the induced electric charge puts the speculation about sizeable electric charge of astrophysical black holes on the firm theoretical ground. This, in turn, provides new perspectives for the detection of a black hole charge in future astrophysical observations.

\begin{acknowledgments}
The work of AK was partially supported by Shota Rustaveli National Science Foundation of Georgia (SRNSFG) through the grant DI-18-335. 
\end{acknowledgments}

\end{document}